\documentclass[]{aa} 
 
\usepackage{graphicx, times, amsfonts}

\newcommand{\less}{\raisebox{-1.1mm}{$\stackrel{<}{\sim}$}}

\begin{document} 
 
%%% referee version 
%\bdouble 
%%%%% MIMIC PRINTED VERSION %%%%%%%
%\rdouble 
 
\title{
The red clump absolute magnitude based on revised Hipparcos parallaxes
}  
 
\author{ 
M.A.T.~Groenewegen 
}

\institute{ 
Koninklijke Sterrenwacht van Belgi\"e, Ringlaan 3, B--1180 Brussels, Belgium \\ \email{marting@oma.be}
} 
 
\date{received: 2008,  accepted:  2008} 
 
\offprints{Martin Groenewegen} 
 
%\authorrunning{M.~ Groenewegen} 
%\titlerunning{LMC and SMC FO Cepheids} 
 
\abstract{Over the past decade the use of the red clump (RC) as distance
  indicator has increased in importance as this evolutionary phase is
  well populated and a good local calibration exists.}
{The absolute calibration of the RC in the $I$ and
  $K$ band is investigated again based on the recently published revised Hipparcos parallaxes.}
{A numerical model is developed that takes the various selection
  criteria and the properties of the Hipparcos catalogue into
  account. The biases involved in applying certain selections are
  estimated with this model.}
{The absolute magnitudes that are found are $M_{\rm I} = -0.22 \pm 0.03$ 
and $M_{\rm K} = -1.54 \pm 0.04$ (on the 2MASS system). The $I$-band value is in good
agreement with previous determinations, the $K$-band value is fainter
than previously quoted, and this seems to be related to a selection
bias whereby \it accurate \rm $K$-magnitudes are only available for relatively few bright stars.  
Applying population corrections to the absolute $K$ magnitude of RC stars
in clusters supports the fainter magnitude scale.}
{}


\maketitle

\section{Introduction} 

Paczy\'nski \& Stanek (1998) were the first to advocate the use of red
clump (RC) stars in the $I$-band as distance indicators. Two
advantages are that this phase of stellar evolution is usually well
populated, and that many RC stars exist with good parallaxes in the
Hipparcos catalogue.

They consider stars with an error in parallax smaller than 10\%, located
in the box 0.8 $< (V-I) <$ 1.25 and $-1.4 < M_{\rm I} <$ 1.2 and with
Hipparcos flag H42 equal to one of ``A,C,E,F,G'', which means that the
star has one or more $I$-band measurements in the Cousins, Johnson, or
Kron-Eggen system.  They find no significant trend of $M_{\rm I}$
on colour and derived $M_{\rm I} = -0.185 \pm 0.016$ for the entire sample. 
They then considered the 228 stars with $d<$70 pc and average distance
of 50 pc to find $M_{\rm I} = -0.192 \pm 0.023$, and $M_{\rm I}=
-0.094 \pm 0.027$ for the rest of the sample at an average distance
of 106 pc. The difference is ascribed to reddening, and when extrapolating
to 0 pc, they arrive at their final result of find $M_{\rm I} = -0.279$
and assign an error bar of 0.088.  In deriving these number, they also
applied a correction for ``a distance bias''.

Stanek \& Garnavich (1998) considered the same sample, and their final
value of $M_{\rm I} = -0.23 \pm 0.03$ is based on the sub sample of
stars within 70 pc and they argue that no correction for reddening is
needed, and they do not apply a correction for selecting the sample on parallax.

Udalski (2000) considered the effect of metallicity. He took the same
cut in parallax error and Hipparcos H42 flags, but also included stars
with flag ``H'' to increase the sample and considered stars with a
metallicity determination from high-resolution spectroscopy by
McWillam (1990). Based on this sample of 284 stars, he finds a weak
dependence on metallicity: $M_{\rm I} = (0.13 \pm 0.07) ($[Fe/H]$
+0.25) + (-0.26 \pm 0.02)$.

Alves (2000) was the first to consider the $K$-band for RC stars as
a standard candle. He considered the 284 stars from Udalski (2000) and
finds that 238 of them have a $K$-magnitude in the Two Micron Sky
Survey (TMSS, Neugebauer \& Leighton 1968; stars are designated by
``IRC'' numbers).  Assuming no reddening and considering stars with
$-2.5 < M_{\rm K} < -0.8$ and all $(V-K)$ colours,  he finds 
$M_{\rm K} = -1.61 \pm 0.03$ and a weak dependence on metallicity 
$M_{\rm K} = (0.57 \pm 0.36) $[Fe/H]$ + (-1.64 \pm 0.07)$.

The RC has gained importance as a distance indicator since the
Paczy\'nski \& Stanek (1998) paper as shown for example by
distance estimates based on the RC to Fornax (Rizzi et al. 2007), M33
(Kim et al. 2002), or IC 1613 (Dolphin et al. 2001).

With the recent publication of the revised Hipparcos parallax data, it
is timely to investigate the absolute magnitude of the RC again.  In
Sect.~2 the revised Hipparcos data is introduced with auxiliary data
used in the analysis. Section~3 describes the numerical model used for
estimating selection biases.  Section~4 describes the results, and
Section~5 presents a brief summary and discussion.

\section{The Data} 
 
The Hipparcos data used are the parallax and error in the parallax
from the re-reduction of the raw data by van Leeuwen (2007).  
For other data ($V$, $I$, Hipparcos flags), the originally published catalogue is
used (ESA 1997).

Using the script submission tool within VizieR (Ochsenbein et
al. 2000) the list of Hipparcos objects was correlated with selected
measurements available, in particular, spectroscopic [Fe/H] values
(available for 2445 stars, and adding 11 stars from Zhao et al. 2001,
64 from Liu et al. 2001, and 178 from Mishenina et al. 2006), the TMSS
catalogue (Neugebauer \& Leighton 1969, available for 3167 stars), and
Str\"omgren photometry (available for 36337 stars). In case of
multiple [Fe/H] determinations the mean was taken.

Independently, the Hipparcos objects are correlated on position with
the DENIS (Epchtein et al. 1999, for 42696 stars) and 2MASS (Skrutskie
et al. 2006, for 117358 stars) NIR catalogues.
The TMSS was the main source of $K$-magnitudes used by Alves (2000).
This is the only source of uniform $K$-magnitudes for bright objects
that often saturate in the DENIS and 2MASS surveys. In the present
study TMSS, DENIS and 2MASS are being used but this requires
transformation to the 2MASS system, which will be the system of choice.
Using 749 objects with a 2MASS photometric quality flag, phqual-flag,
of ``A'' in the $K$-band and that not saturate in DENIS $I$ (implying
$I >10$) it is found that DENIS $K$ - 2MASS $K$ = $-0.018 \pm 0.066$
with no colour term on $(I-K)_{\rm DENIS}$.
There are no stars in common between the TMSS survey and non-saturated
DENIS or 2MASS objects with phqual-flags of ``A'' or ``B'' in $K$. The
transformation to the 2MASS system is done indirectly via the SAAO
Carter (1990) system. The advantage over a comparison to the Koornneef
(1983) system, which was used by Alves (2000), is that the colour term
and offset with respect to the 2MASS system are smaller (Carpenter 2001, and
updated on the IPAC website\footnote{
www.ipac.caltech.edu/2mass/releases/allsky/doc/sec6\_4b.html}).

Using 62 TMSS stars that are listed in Carter (1990) and that have both 
an IRC $K$ {\it and} $I$ magnitude available, the following relations are derived:
$(J-K)_{\rm Carter} = 0.484 (I-K)_{\rm IRC} -0.103$ with a dispersion of 0.04 mag, and eliminating 5 outliers,
%$(J-K)_{\rm Carter} = 0.443 (I-K)_{\rm IRC} -0.029$ with a dispersion of 0.03 mag, and eliminating 2 outliers,
$K_{\rm Carter} - K_{\rm IRC} = -0.011$ with no significant colour term, and with a dispersion of 0.04 mag.
%$K_{\rm Carter} - K_{\rm IRC} = 0.140 (I-K)_{\rm IRC} -0.234$ with a dispersion of 0.03 mag.
The transformation from the Carter to the 2MASS system follows the relation on the quoted IPAC website: 
$K_{\rm 2mass} = K_{\rm Carter} -0.024 + 0.017 (J-K)_{\rm Carter}$.

The adopted $K$-band magnitude is, in order of preference, 
(1) the 2MASS value, but only when phqual is ``A'', 
(2) The DENIS value (for non-saturated stars with $I_{\rm denis} > 10$) corrected by +0.018 mag, 
(3) the TMASS/IRC value, transformed to the 2MASS system as described above. 
If no $I_{\rm IRC}$ is available a mean colour for RC stars of $(I-K)_{\rm IRC} = 1.55$ is used in the transformation.

Of all Hipparcos stars that have a $K$-magnitude assigned, 3116 are based on TMSS, 17 on Denis and 103827 on 2MASS.

%A plot of ($K_{\rm IRC} - K_{\rm 2MASS}$) versus $(I-K)_{\rm IRC}$ colour 
%using almost 2500 stars shows no significant slope (0.004 $\pm$ 0.004) or off-set (0.018 $\pm$ 0.011).
%

\section{The model} 
 
In this section a model is described to construct synthetic samples of
stars and apply various selection criteria in order to generate
samples of stars that closely resemble in nature the observed data.
The model is follows largely Groenewegen \& Oudmaijer (2000) but is
based on the properties of the revised Hipparcos catalogue.

\subsection{Galactic model}

The coordinate system used is cylindrical coordinates centred on the
Galactic centre. The galactic distribution is assumed to be a double
exponential disk with a scale height $H$ in the $z$-direction (the
coordinate perpendicular to the galactic plane), and a scale length
$R_{\rm GC}$ in the galacto-centric direction.  The distance to the
Galactic centre is taken to be 7800 pc (Zucker et al. 2006).

\subsection{Luminosity function}

The luminosity function can be arbitrary but, in the case of simulating
RC stars, is assumed to be a Gaussian. In $I$ with a mean $M_{\rm I} = -0.27$ 
and $\sigma$ of 0.20.  The $(V-I)$ colours is assumed to be a Gaussian
with mean 0.98 and $\sigma$ = 0.085.
In $K$ the Gaussian has a mean $M_{\rm K} = -1.60$ and $\sigma$ of 0.22, 
while the $(V-K)$ colours is taken as a Gaussian with mean 2.32 and $\sigma$ of 0.21.

\subsection{Reddening}
\label{S-red}

Visual extinction in the simulation and to de-redden the observations
is based on several 3-dimensional reddening models available in the literature.
Marshall et al. (2006) presents reddening in the $K$-band with
15$\arcmin$ sampling along 64~000 lines-of-sight in the direction
$\mid l \mid \le 100\deg$ and $\mid b \mid \le 10\deg$ based on 2MASS data.
Drimmel et al. (2003) presents a reddening model based on the dust
distribution model of Drimmel \& Spergel (2001), which is based on COBE/DIRBE data.
Arenou et al. (1992) presents a reddening model based on a comparison of
observed $B,V$ photometry with predicted photometry from a star's spectral type.
Finally, a simple law following Parenago (1940) was considered:
\begin{equation}
      A_{\rm V}= 0.097/ \mid \sin b \mid  (1.0 - \exp(-0.0111 \; d \; \mid \sin b \mid))
\end{equation}
The Marshall et al. results have been taken as reference model, and
their reddenings were transformed to $A_{\rm V}$ as $A_{\rm K}/0.12$.  

However, this model is only available over a limited range in galactic
coordinates, therefore a Monte Carlo simulation was made to randomly
generate galactic coordinates and distances assuming a constant number
density of stars within 3 kpc distance of the Sun. If the galactic
coordinates were within the range of applicability of the Marshall et
al. model, the visual extinction from the other models was determined.
In the end, using several 1000 positions, the average and dispersion
in the various ratios between extinction models were calculated, and
the results are in Table~\ref{Tab-red}, in the sense model listed in
the row divided by model listed in the column.

Although the dispersion in all the ratios is quite large, they mostly
agree in the mean. It only seems that the Arenou et al. model gives
higher extinction than the Drimmel et al. and Parenago model.  The
finally adopted visual extinction was the average of the models by 
Parenago, Drimmel et al., and Arenou et al. multiplied by 0.84.

\begin{table} 
\caption{Comparison of different reddening models} 
\begin{tabular}{lcccccc} \hline \hline 
                 & Drimmel et al.  & Parenago        & Arenou et al.   \\ \hline
Marshall et al. & 1.01 $\pm$ 0.82 & 1.00 $\pm$ 0.88 & 0.99 $\pm$ 0.76 \\
Arenou et al.    & 1.20 $\pm$ 0.62 & 1.15 $\pm$ 0.55 & -               \\
Parenago         & 1.06 $\pm$ 0.24 & -               & -               \\
\hline 
\end{tabular} 
\label{Tab-red}
\end{table}

\subsection{Properties of Hipparcos data}

The adopted completeness function in $Hp$ magnitude is
\begin{equation}
 \left(1.0 + \exp ((Hp - 8.695)/0.5206) \right)^ {-1.194}.
\end{equation}
This was derived by comparing the observed magnitude distribution to
that predicted with the TRILEGAL Galactic model (Girardi et al. 2005).

The completeness functions quoted below are derived by inspecting the
ratio of stars that fulfil a particular selection to all stars, as a
function of $Hp$ magnitude.
The completeness function of stars with $I$-band flags ``A+C+E+F+G'' is approximated as
\begin{equation}
\begin{tabular}{ll}
 0.90                   &  Hp $\le$ 5 \\
 0.50 - 0.45 (Hp - 5)/2 &  5 $<$ Hp $<$ 7 \\
 0.05                   &  Hp $\ge$ 7.  \\
\end{tabular} 
\end{equation}
If stars with $I$-band flag ``H'' are also included, this becomes
\begin{equation}
\begin{tabular}{ll}
 0.92                    & Hp $\le$ 5 \\
 0.60 - 0.40 (Hp - 5)/4  & 5 $<$ Hp $<$ 9 \\
 0.20                    & Hp $\ge$ 9.
\end{tabular} 
\end{equation}

As outlined above. $K$-band magnitudes were to Hipparcos
stars based on high-quality 2MASS and IRC data. The completeness
function is approximated as
\begin{equation}
\begin{tabular}{ll}
 0.60                      & Hp $\le$ 5 \\
 0.69                      & 5 $<$ Hp $\le$ 7 \\
 0.69 + 0.30 (Hp - 7)/1.5  & 7 $<$ Hp $<$ 8.5 \\
 0.99                      & Hp $\ge$ 8.5.
\end{tabular} 
\end{equation}
For fainter objects, a (reliable) $K$-magnitude is available for almost
all objects because of the all-sky nature of the 2MASS survey, but for
brighter stars this is clearly not the case.

As outlined above, spectroscopic [Fe/H] values have been retrieved
using Vizier and recent literature. The completeness function is
approximated as
\begin{equation}
\begin{tabular}{ll}
 0.55                   & Hp $\le$ 4 \\
 0.55 - 0.53 (Hp - 4)/3 & 4 $<$ Hp $<$ 7 \\
 0.02                   & Hp $\ge$ 7 \\
 0.0                    & Hp $\ge$ 8. \\
\end{tabular} 
\end{equation}
The error on the $Hp$ magnitude is small but has nevertheless been
taken into account. The error is Gaussian distributed with the sigma value
\begin{displaymath}
 \log \sigma_{\rm Hp} = 0.290 Hp -5.510 \hspace{10mm} {\rm Hp > 11}
\end{displaymath}
\begin{equation}
 \log \sigma_{\rm Hp} = 0.177 Hp -4.265 \hspace{10mm} {\rm Hp \le 11}.
\end{equation}

As explained in the explanatory supplement to the Hipparcos catalogue, stars
with an ecliptic latitude more than 47 degree were observed more often, 
and this resulted in lower parallax errors on average. 
When $\beta >$ 47\degr\ the error on the parallax (in mas) is in the mean
\begin{displaymath}
 \log \sigma_{\pi}= 0.1610 Hp -1.540,
\end{displaymath}
with dispersion  0.075, and else
\begin{equation}
 \log \sigma_{\pi}= 0.1573 Hp -1.368,
\end{equation}
with dispersion 0.063.
The minimum error on the parallax is
\begin{equation}
(\log \sigma_{\pi})_{\rm min}= 0.1610 Hp -1.7
\end{equation}
In the simulation a parallax error is generated using the mean
relation and the Gaussian dispersion. If this parallax error is above
the minimum allowed value the star is retained.

The ``measured'' parallax is based on the true parallax (from the true
distance) and the Gaussian distributed parallax error.

\subsection{The model in practice}

To begin with, a number of input parameters need to be set:
\begin{itemize}

\item scale length and scale height of the population,

\item mean magnitude and dispersion in the luminosity function in the
  reference magnitude, $\lambda$ (i.e. $I$ or $K$ in the present
  paper),

\item mean magnitude and dispersion in the luminosity function in the ($V-\lambda$)-colour,

\item $A_{\lambda}/A_{\rm V}$ value, and 

\item the number of simulations, and number of stars per simulation.

\end{itemize}

\noindent
The consecutive steps in the model are the following:
\begin{itemize}

\item Three random numbers are drawn to select the distance to the
Galactic plane, the distance to the Galactic centre and a random angle
$\phi$ between 0 and 2$\pi$ in the Galactic plane centred on the
Galactic centre. From this the distance $d$ to the Sun is calculated. 
The galactic coordinates and ecliptic latitude is determined.

From the properties of the observed sample (e.g. from the observed
parallax, or a photometrically determined parallax), it is already
possible to eliminate stars beyond a certain limiting distance.

\item The reddening in the $V$-band and in the reference magnitude is determined.

\item The true absolute magnitude in $V$ and in the reference magnitude is determined.

\item The true $Hp$ magnitude is determined, from the distance, 
$M_{\rm v}$, $A_{\rm V}$, and the fact that $(Hp - V)$ is typically
+0.1 (vol. 1, sect 1.3, p. 59 of Hipparcos explanatory catalogue).

\item The error in the $Hp$ magnitude is determined and the observed $Hp$ determined.

\item The completeness function in $Hp$ is determined, and a random
  number is generated to evaluate if the star is ``observed''.

\item Depending on the selection of the sample, other completeness functions are evaluated and applied.

\item If the star is still ``in'', Eqs.~(8-9) are evaluated to
determine the parallax errors, and the observed parallax.

\item Depending on the selection of the sample, additional selection
criteria on parallax or (relative) parallax error are applied.

\item The observed magnitude in the reference colour is determined.

\item The output of the model typically is the ``observed'' parallax,
error in the parallax, true distance, photometric parallax based on
the adopted absolute mean magnitude, $Hp$, observed reference magnitude, 
$A_{\rm V}$ and galactic coordinates.

\end{itemize}

\section{The red clump}

Table~\ref{Tab-Res} lists the observed absolute magnitudes depending
on various selections. These selections largely follow the selections
made by previous works as quoted in the Introduction. In all cases, 
only stars where Hipparcos flag H30 (the goodness-of-fit) is $<$5,
$\pi >0$, and single stars (isoln = 5 as listed in the revised
Hipparcos catalogue) are considered.

Following earlier work, stars are selected inside the box (all colours
and magnitudes are understood to be de-reddened according to the model
described in Sect.~\ref{S-red}) 0.75 $< (V-I) <$ 1.25, $-1.6 < M_{\rm  I} <$ 1.2.
The data is binned in absolute magnitude and then fitted with the
formula introduced by Paczy\'nski \& Stanek (1998):

\begin{displaymath}
 N= a_1 + a_2 (m- m_0) + a_3 (m -m_0)^2 
%                     + a_4 / ( \sqrt{2 \pi} \sigma)  \; \exp (-\frac{1}{2} ( (m -m_0)/\sigma)^2)
\end{displaymath}
\begin{equation}
% N= a_1 + a_2 (m- m_0) + a_3 (m -m_0)^2 \\
 \hspace{20mm}     + \; \frac{a_4}{ \sqrt{2 \pi \sigma_{\rm RC}^2} } \; \exp (-\frac{1}{2} ( (m -m_0)/\sigma_{\rm RC})^2).
\label{Eq-rc}
\end{equation}

\begin{figure}

\begin{minipage}{0.4\textwidth}
\resizebox{\hsize}{!}{\includegraphics{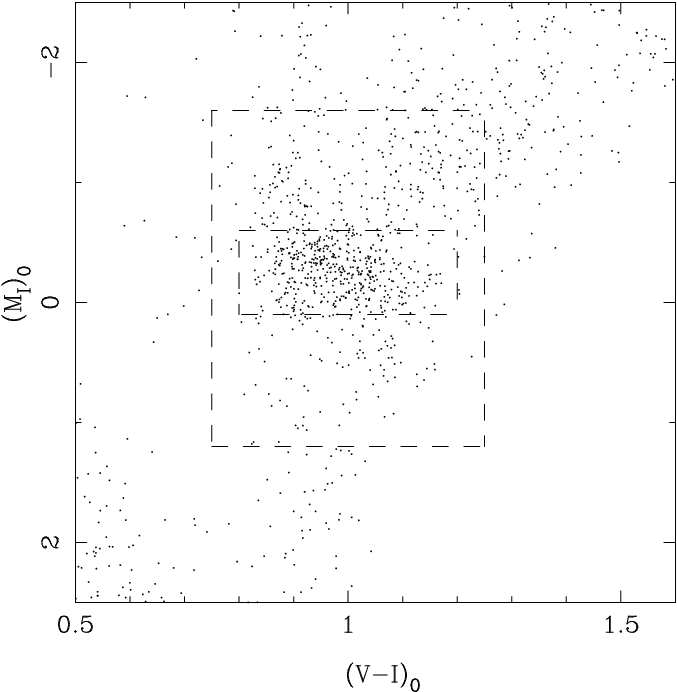}}
\end{minipage}

\begin{minipage}{0.4\textwidth}
\resizebox{\hsize}{!}{\includegraphics{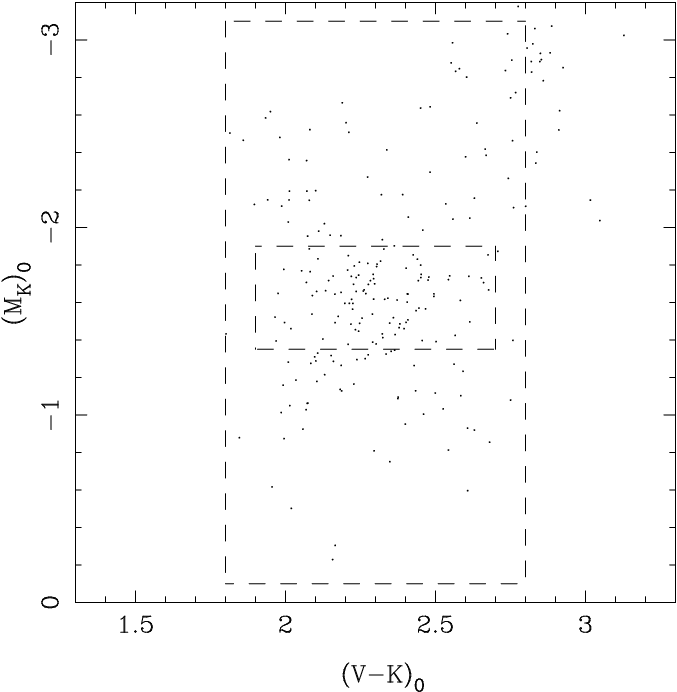}}
\end{minipage}

\caption[]{
Colour-magnitude diagrams for model 3. The outer box is used to select
the data to fit Eq.~\ref{Eq-rc} to. The inner box is used to fit the dependence
of the absolute magnitude against colour or metallicity.
}
\label{Fig-Box}
\end{figure}

\begin{figure}

\begin{minipage}{0.4\textwidth}
\resizebox{\hsize}{!}{\includegraphics{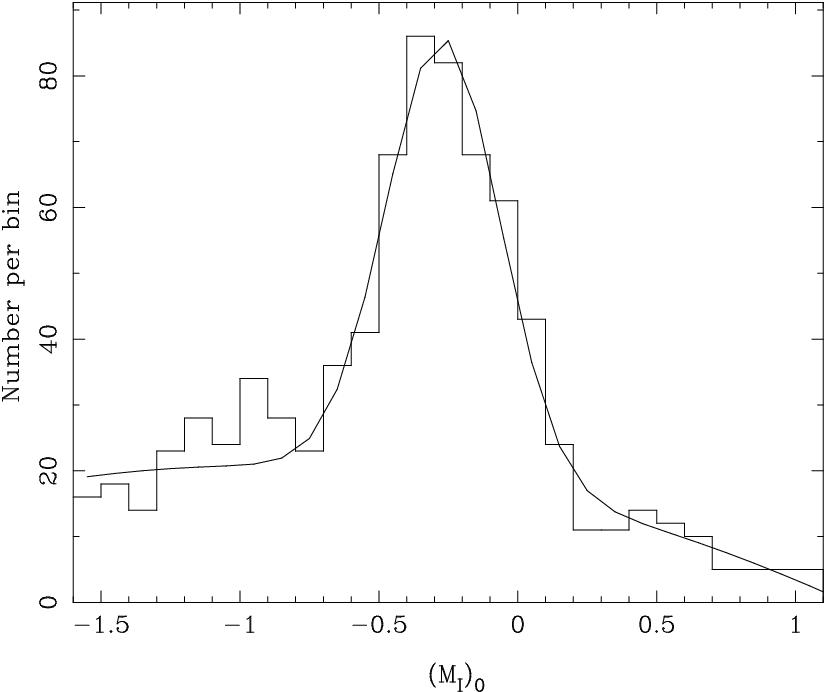}}
\end{minipage}

\begin{minipage}{0.4\textwidth}
\resizebox{\hsize}{!}{\includegraphics{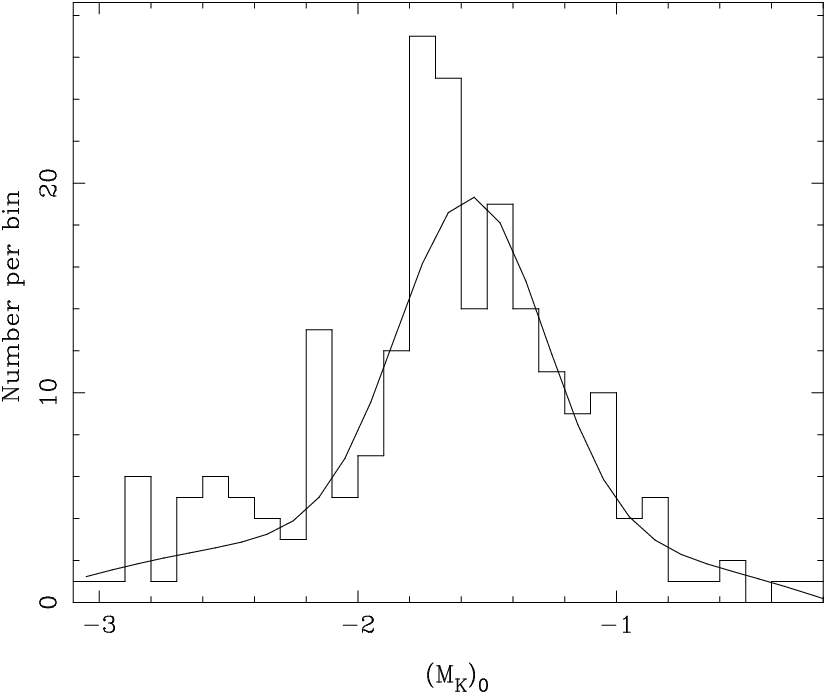}}
\end{minipage}

\caption[]{
The fit of Eq.~\ref{Eq-rc} to the data in the $I$-band (upper panel), and $K$-band for model 3.
}
\label{Fig-Fit}
\end{figure}

The results in the $K$-band are based on a subset of the selected
stars in the $I$-band. In this case Eq.~\ref{Eq-rc} is fitted to the
stars inside the box defined by 1.8 $< (V-K) <$ 2.8 and $-3.1 < M_{\rm  K} < -0.1$.
The location of the boxes in a colour-magnitude diagram are illustrated in Fig.~\ref{Fig-Box}.
An illustration of the fit of Eq.~\ref{Eq-rc} to the data is shown in Fig.~\ref{Fig-Fit}.

Several of the entries in Table~\ref{Tab-Res} are put between
parentheses, to indicate a lower quality. In the $I$-band this
refers to selections that include lower-quality $(V-I)$ data. In $K$
this refers to model 5 that includes too few stars for an
accurate fit of the Gaussian.

As the selection box and Eq~\ref{Eq-rc} are designed to include a
``background'' of stars, the fit of the absolute magnitude against
colour or metallicity is made for stars that are located inside a
smaller box. Based on the fit with Eq.~\ref{Eq-rc} and the coefficients
$a_4,m_0,\sigma_{\rm RC}$, the number of red clump stars can be estimated. 
The location of the smaller box is chosen such that it approximately
includes this number of stars.  In the $I$-band this is 0.8 $< (V-I) <$ 1.2 
and $-0.6 < M_{\rm I} <$ 0.1. In the $K$-band this is 1.9 $< (V-K) <$ 2.7 
and $-1.9 < M_{\rm K} < -1.35$.

First the dependence on colour and metallicity is investigated,
fitting a linear relation.  When $M_{\rm I}$ is fitted against $(V-I)$
colour, the slope is $0.27 \pm 0.11$, $0.10 \pm 0.19$, and $-0.02 \pm 0.22$, 
respectively, when all, ``ACEFGH'' and ``ACEFG'' Hipparcos
$I$-band flags are considered.
For these three selections, the slope is $0.08 \pm 0.06$, $0.06 \pm 0.07$,  
and $0.09 \pm 0.08$ when $M_{\rm I}$ is fitted against [Fe/H]. 
The three determinations are consistent, as one would expect since any dependence
on metallicity should not depend on the origin of the $I$-band data.
No significant dependence on colour is found, in accordance with
Paczy\'nski \& Stanek (1998). Also the dependence on metallicity is marginal.

When $M_{\rm K}$ is fitted against $(V-K)$ colour, the slope is 
$-0.15 \pm 0.07$, $-0.15 \pm 0.08$, and $-0.13 \pm 0.09$, respectively, 
when all, ``ACEFGH'' and ``ACEFG'' Hipparcos $I$-band flags are considered.
Fitted against [Fe/H], the slopes become, respectively, 
$0.02 \pm 0.07$, $-0.03 \pm 0.08$, and $-0.04 \pm 0.09$. 
In $K$ there is therefore no evidence of any metallicity dependence
and only a marginal one on $(V-K)$ colour.

\begin{figure}

\begin{minipage}{0.49\textwidth}
\resizebox{\hsize}{!}{\includegraphics{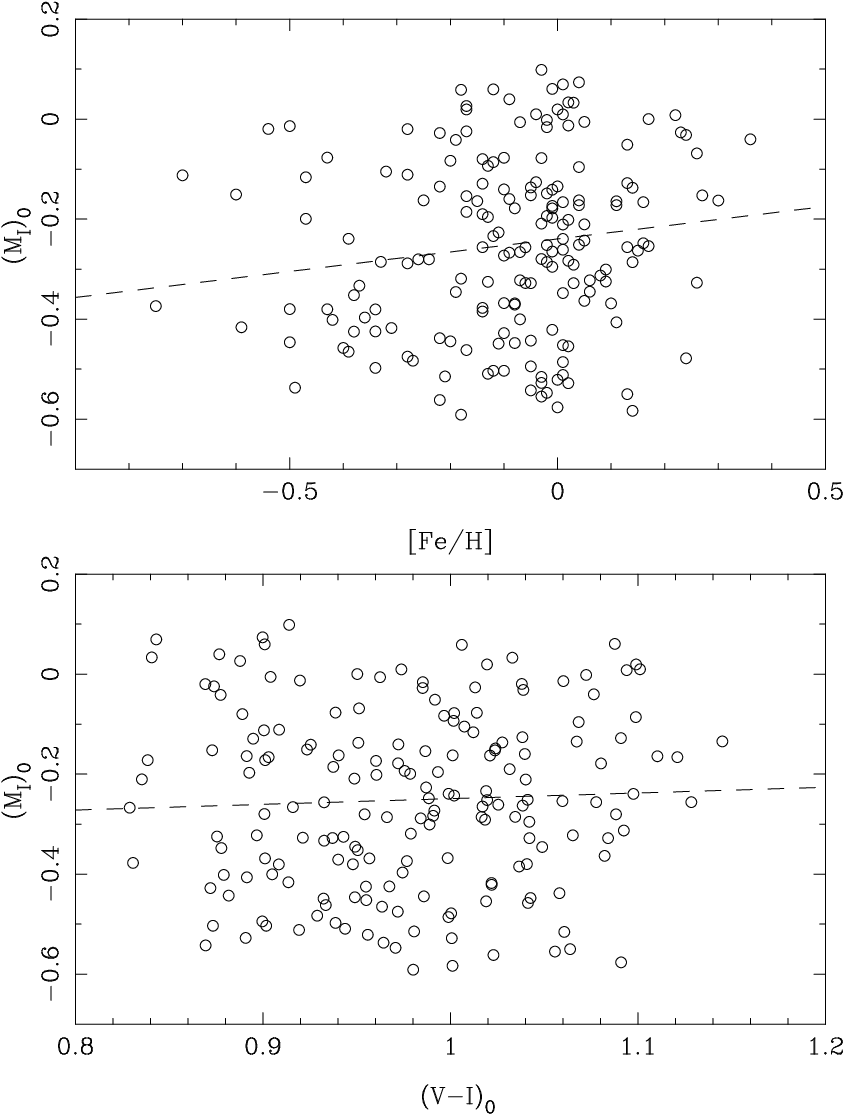}}
\end{minipage}

\caption[]{
Fit of $M_{\rm I}$ versus metallicity and $(V-I)$ colour for model 8, with best fits indicated by the dashed line. 
The slope of the fit against colour is not significant.
}
\label{Fig-MI}
\end{figure}

\begin{figure}

\begin{minipage}{0.49\textwidth}
\resizebox{\hsize}{!}{\includegraphics{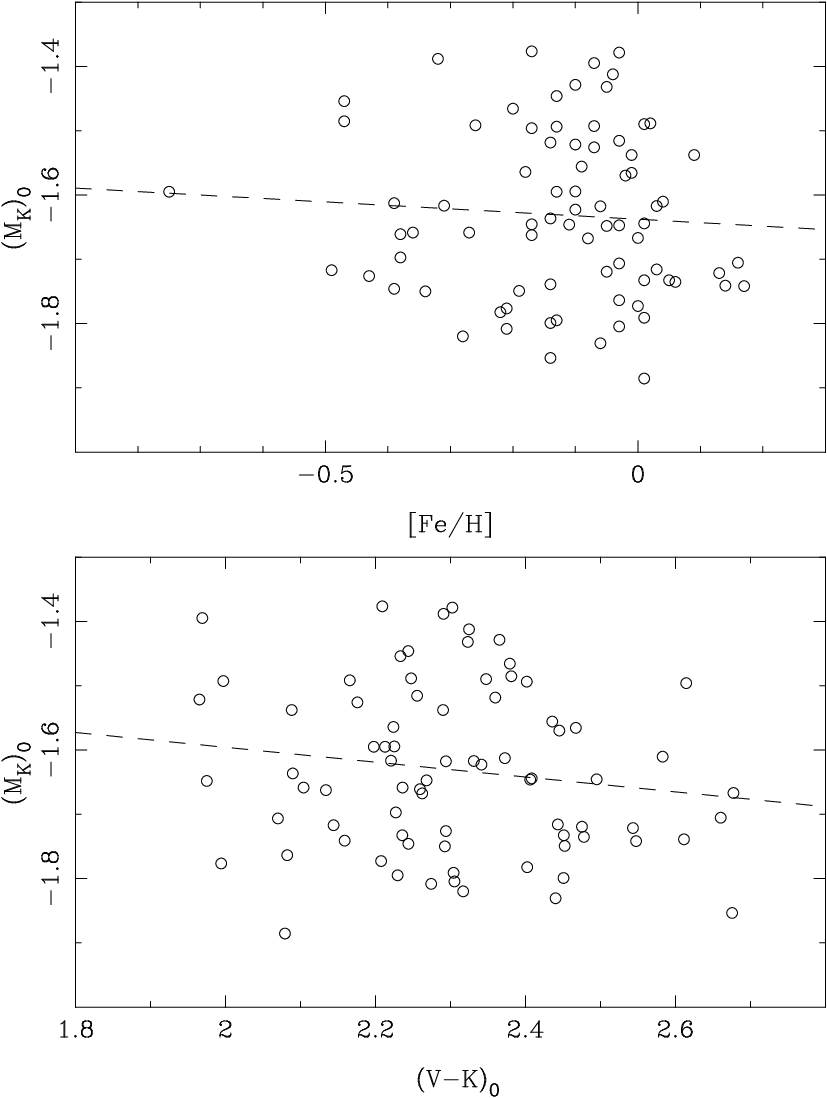}}
\end{minipage}

\caption[]{
Fit of $M_{\rm K}$ versus metallicity and $(V-K)$ colour for model 8, with best fits indicated by the dashed line. 
The slope of the fit against metallicity is not significant.
}
\label{Fig-MK}
\end{figure}

\begin{table*} 
\caption{Results based on different sample selections} 
\begin{tabular}{cllcrccrccc} \hline \hline 
Model & $\sigma_{\pi}/\pi$ & $I$-flags & [Fe/H] & N &   $M_{\rm I}$  & $\sigma_{\rm RC,I}$ &    N & $     M_{\rm K}$     & $\sigma_{\rm RC,K}$  \\
\hline
%                                                                                                                                          nBG nRC nBox ; nBG nRC
1& $<0.1$             & all    & No  & 3395 & (-0.206 $\pm$ 0.011) & 0.249 $\pm 0.013$ & 1593 &  -1.398 $\pm$ 0.017  & 0.369 $\pm$ 0.021 \\
2& $<0.1$             & ACEFGH & No  & 1411 & (-0.245 $\pm$ 0.014) & 0.217 $\pm 0.016$ &  436 &  -1.534 $\pm$ 0.029  & 0.277 $\pm$ 0.036 \\ %780+631 677;180+228 Ok
3& $<0.1$             & ACEFG  & No  &  795 &  -0.269 $\pm$ 0.016  & 0.205 $\pm 0.018$ &  236 &  -1.582 $\pm$ 0.035  & 0.271 $\pm$ 0.043 \\ %410+385 414; 80+134 Pff
4& $<0.1$, $\pi>$14.3 & ACEFG  & No  &  182 &  -0.259 $\pm$ 0.023  & 0.193 $\pm 0.026$ &  116 &  -1.624 $\pm$ 0.025  & 0.146 $\pm$ 0.025 \\ % 60+122 101; 50+ 66 UN
5& $<0.1$, $\pi>$20.0 & ACEFG  & No  &   86 &  -0.232 $\pm$ 0.033  & 0.184 $\pm 0.035$ &   60 & (-1.628 $\pm$ 0.033) & 0.091 $\pm$ 0.029 \\ % 30+ 56  51; 30+ 30 UN

%6& $<0.1$             & all    & Yes &  357 & (-0.259 $\pm$ 0.020) & 0.179 $\pm 0.020$ &  213 &  -1.630 $\pm$ 0.025  & 0.146 $\pm$ 0.024 \\ %170+187 189;110+103 Ok
%7& $<0.1$             & ACEFGH & Yes &  294 & (-0.271 $\pm$ 0.022) & 0.190 $\pm 0.023$ &  187 &  -1.629 $\pm$ 0.027  & 0.158 $\pm$ 0.027 \\ %130+164 159; 90+ 97 Ok
%8& $<0.1$             & ACEFG  & Yes &  225 &  -0.297 $\pm$ 0.027  & 0.205 $\pm 0.029$ &  149 &  -1.624 $\pm$ 0.031  & 0.153 $\pm$ 0.031 \\ %100+125 123; 70+ 79 Pff

6& $<0.1$             & all    & Yes &  487 & (-0.217 $\pm$ 0.016) & 0.185 $\pm 0.017$ &  245 &  -1.640 $\pm$ 0.022  & 0.149 $\pm$ 0.022 \\ %200+287 265;110+117 
7& $<0.1$             & ACEFGH & Yes &  377 & (-0.232 $\pm$ 0.018) & 0.199 $\pm 0.019$ &  210 &  -1.642 $\pm$ 0.024  & 0.153 $\pm$ 0.023 \\ %140+237 215;100+ 97 
8& $<0.1$             & ACEFG  & Yes &  285 &  -0.249 $\pm$ 0.021  & 0.206 $\pm 0.023$ &  156 &  -1.650 $\pm$ 0.025  & 0.140 $\pm$ 0.025 \\ %110+175 165; 80+ 76 

9& $<0.05$            & ACEFGH & No  &  716 & (-0.240 $\pm$ 0.014) & 0.186 $\pm 0.014$ &  226 &  -1.651 $\pm$ 0.024  & 0.150 $\pm$ 0.024 \\ %340+376 386;110+ 96 Ok

\hline 

\hline 
\end{tabular} 
\label{Tab-Res}
\end{table*}

For the selections in Table~\ref{Tab-Res}, the results of the
simulations are listed in Table~\ref{Tab-Sim}.  As mentioned in
Sect.~3.2, the input values are $M_{\rm I} = -0.27$ with $\sigma$ = 0.20, 
and $M_{\rm K} = -1.60$ with $\sigma$ = 0.22.
One can observe that the bias can be up to 0.07 mag in $I$ depending
on the selection bias. The fifth column in Table~\ref{Tab-Sim} lists
the ``true'' value, which is the observed value from Table~\ref{Tab-Res}, 
corrected by the bias, estimated from the difference between the input
value and the output value from the simulation.

The average of the 4 available models or the one with smallest correction
(model 5) gives the identical result: $M_{\rm I} = -0.22 \pm 0.03$.
In $K$ the biases are larger, up to 0.1 mag, and the average of 7
estimates is $M_{\rm K} = -1.54 \pm 0.04$.

\begin{table*} 
\caption{Results of the simulations} 
\begin{tabular}{ccccccc} \hline \hline 
Model & $M_{\rm I}$    & $\sigma_{\rm RC,I}$ &     $M_{\rm K}$     & $\sigma_{\rm RC,K}$    &   $M_{\rm I}$ (true) & $M_{\rm K}$ (true) \\
\hline

2 & (-0.290 $\pm$ 0.010) & 0.240 $\pm 0.007$ &  -1.648 $\pm$ 0.018  & 0.251 $\pm$ 0.015 &          -         & $-1.486 \pm$ 0.034 \\ 
3 &  -0.327 $\pm$ 0.012  & 0.222 $\pm 0.010$ &  -1.697 $\pm$ 0.022  & 0.238 $\pm$ 0.021 & $-0.212 \pm$ 0.020 & $-1.485 \pm$ 0.041 \\ 
4 &  -0.290 $\pm$ 0.023  & 0.190 $\pm 0.021$ &  -1.625 $\pm$ 0.035  & 0.202 $\pm$ 0.025 & $-0.239 \pm$ 0.033 & $-1.599 \pm$ 0.043 \\ 
5 &  -0.264 $\pm$ 0.053  & 0.214 $\pm 0.040$ & (-1.598 $\pm$ 0.110) & 0.242 $\pm$ 0.099 & $-0.238 \pm$ 0.062 &          -         \\ 
%6 & (-0.330 $\pm$ 0.018) & 0.219 $\pm 0.014$ &  -1.701 $\pm$ 0.028  & 0.236 $\pm$ 0.024 &           -        & $-1.529 \pm$ 0.038 \\ 
%7 & (-0.335 $\pm$ 0.020) & 0.210 $\pm 0.018$ &  -1.703 $\pm$ 0.031  & 0.236 $\pm$ 0.026 &           -        & $-1.526 \pm$ 0.041 \\ 
%8 &  -0.339 $\pm$ 0.022  & 0.209 $\pm 0.019$ &  -1.717 $\pm$ 0.035  & 0.227 $\pm$ 0.031 & $-0.228 \pm$ 0.035 & $-1.507 \pm$ 0.047 \\ 
6 & (-0.333 $\pm$ 0.015) & 0.218 $\pm 0.011$ &  -1.702 $\pm$ 0.025  & 0.234 $\pm$ 0.021 &           -        & $-1.538 \pm$ 0.033 \\ 
7 & (-0.337 $\pm$ 0.018) & 0.208 $\pm 0.015$ &  -1.707 $\pm$ 0.027  & 0.233 $\pm$ 0.029 &           -        & $-1.535 \pm$ 0.036 \\ 
8 &  -0.330 $\pm$ 0.021  & 0.217 $\pm 0.018$ &  -1.715 $\pm$ 0.030  & 0.231 $\pm$ 0.029 & $-0.189 \pm$ 0.034 & $-1.535 \pm$ 0.039 \\ 

9 & (-0.306 $\pm$ 0.011) & 0.210 $\pm 0.009$ &  -1.664 $\pm$ 0.028  & 0.228 $\pm$ 0.021 &           -        & $-1.587 \pm$ 0.037 \\

\hline 

\hline 
\end{tabular} 
\label{Tab-Sim}
\end{table*}

\section{Summary and discussion} 

Using revised Hipparcos parallaxes and extensive numerical simulations
that take into account de properties of the Hipparcos catalogue in terms
of completeness and parallax error, and also taking into account the
selection functions that are involved in selecting stars with certain
properties ($K$-band, high-quality $I$-band, spectroscopic metallicities), 
the absolute magnitude of the red clump in $I$ and $K$ is investigated.

In the $I$-band, the value of $M_{\rm I} = -0.22 \pm 0.03$ essentially 
agrees with the value quoted by Stanek \& Garnavich (1998). 
There is no dependence of $M_{\rm I}$ on $(V-I)$ colour, and
the dependence on metallicity is marginal: $(0.08 \pm 0.07)$ ([Fe/H] + 0.15).

In the $K$-band, the influence of selection biases appears stronger and
the value derived $M_{\rm K} = -1.54 \pm 0.04$ differs significantly
from the value originally derived by Alves (2000), which is almost 0.1 mag brighter, 
when his value of $-1.61$ on the Bessell \& Brett (1988) system is
converted to the 2MASS system (Carpenter 2001).  There is no
dependence of $M_{\rm K}$ on metallicity, and the dependence on colour
is weak: $(-0.15 \pm 0.07)$ ($(V-K)_0 - 2.32$).

The RC stars in clusters have also been used to derive $M_{\rm K}$,
originally by Grocholski \& Sarajedini (2002). Absolute magnitudes are
derived by combining the observed mean magnitude in a box in colour
and magnitude with a reddening estimate and main-sequence fitting
distances.  Grocholski \& Sarajedini (2002) used the second
incremental data release of 2MASS and 14 clusters to find an absolute
$K$-magnitude that is in agreement with the value in Alves (2000).
Using the All-Sky data release and increasing the sample to 24
clusters, Van Helshoecht \& Groenewegen (2007; hereafter vHG) find a
value fainter by 0.05 mag, or $M_{\rm K} = -1.57 \pm 0.05$ on the
Bessell \& Brett system.  Both Grocholski \& Sarajedini (2002) and vHG
used averages over the cluster sample to arrive at the quoted means.

To compare these values to the absolute magnitude of the local
Hipparcos sample, a ``population correction'' has to be made, to
account for the difference in metallicity and age of the RC
population.  The calculation of this correction is outlined and
tabulated in Girardi \& Salaris (2001) and Salaris \& Girardi
(2002). In the $K$-band, Salaris \& Girardi (2002) show that the
correction is a strong function of age for ages below 3 Gyr.  The top
panel in their Fig.~3 shows that the correction is well behaved for
ages above about 4 Gyr, and for each metallicity a linear function was fitted.

From vHG the five clusters older that 4 Gyr were taken and the
population correction was determined by linear interpolation in age
and [Fe/H] in the results in Salaris \& Girardi.  The error in $\Delta
M_{\rm K}$ is based on a 1 Gyr error in age and 0.1 dex in
metallicity.  The results are listed in Table~\ref{Tab-Clusters},
which lists the age and metallicity (see vHG for the references for
age and metallicity determination), the $M_{\rm K}$ value and its
error that takes into account the error in reddening and the assumed
distance to the cluster (from vHG), the population correction with
error, and the corrected value.
The weighted mean of the corrected $M_{\rm K}$ values is $-1.39 \pm$ 0.06 
with a dispersion of 0.2 mag.  Correcting to the 2MASS system
(Carpenter 2001) this becomes $-1.43 \pm$ 0.06.  Although the
dispersion is large, the mean value is also fainter than the straight
mean for the cluster sample and agrees within 2-sigma with the
determination from the Hipparcos sample.

To settle the issue on the importance of the bias and the absolute
$K$-magnitude of RC stars would require accurate NIR magnitudes of a
100 to a few hundred (cf. Table~2) bright ($K \less 5$) RC stars. 
Given the brightness, this represents a challenge to modern
instrumentation because of saturation.

The absolute magnitudes derived here are in better agreement with theory than previously.  
Using a plausible star formation history for the solar neighbourhood, Salaris \& Girardi (2002) derived absolute magnitudes of $M_{\rm I}= -0.17$ 
and $M_{\rm K}= -1.584$ (on the 2MASS system), which they compared to the Hipparcos-based results quoted in Alves et al. (2002) of $M_{\rm I}= -0.26 \pm 0.03$ and 
$M_{\rm K}= -1.644 \pm 0.03$ (on the 2MASS system), which indicates differences at 
the 2-3 sigma level.  The new results of $M_{\rm I}= -0.22 \pm 0.03$ and 
$M_{\rm K}= -1.54 \pm 0.04$ agree with theory at the 1-2 sigma level.

\begin{table*} 
\caption{The RC in old clusters } 
\begin{tabular}{lcccccc} \hline \hline 
Name & $\log t$ & [Fe/H] &     $M_{\rm K}$   & $\Delta M_{\rm K}$ &  $M_{\rm K}$ (true) \\
\hline

Be 39    & 9.90 & -0.15 & $-1.56 \pm 0.11 $ & $-0.058 \pm 0.03$ & $-1.61 \pm 0.12$ \\ 
NGC 188  & 9.63 & -0.12 & $-1.36 \pm 0.11 $ & $+0.057 \pm 0.03$ & $-1.30 \pm 0.12$ \\ 
NGC 6791 & 9.64 & +0.40 & $-1.39 \pm 0.11 $ & $+0.223 \pm 0.05$ & $-1.17 \pm 0.12$ \\ 
47 Tuc   &10.08 & -0.70 & $-1.34 \pm 0.21 $ & $-0.360 \pm 0.11$ & $-1.70 \pm 0.24$ \\ 
NGC 362  &10.08 & -1.15 & $-0.81 \pm 0.24 $ & $-0.680 \pm 0.11$ & $-1.49 \pm 0.26$ \\

\hline 
\end{tabular} 
\label{Tab-Clusters}
\end{table*}

Finally, some implications for existing distance determinations using the RC are discussed.  
The derived value of the absolute magnitude in the $I$-band of $M_{\rm I} = -0.22 \pm 0.03$ 
is not very different from previously adopted values in the
literature of $-0.23$ or $-0.26$, so the impact on derived distances is not so great.

The distance to the LMC quoted by Pietrzy\'nski \& Gieren (2002) of
18.501 $\pm$ 0.049 (random+systematic error and assuming no population
correction) based on $JK$ observations of two fields in the bar would
become 18.35 $\pm$ 0.05 for an assumed population correction of
$-0.03$ (Salaris \& Girardi 2002).

Babusiaux \& Gilmore (2005) obtained infrared data on some fields in
the direction of the Galactic bulge.  Their distance scale is
tied to one field for which they obtain $M_{\rm K} = -1.72$ for an
assumed distance of 8 kpc. Taking $M_{\rm K} = -1.54$
would result in a distance of 7.4 kpc ($\pm$0.2 kpc judging from the
information in their Table~2). Adding the error in $M_{\rm K}$ results in a total error of 0.3 kpc.

Using similar type of observations Nishiyama et al. (2006) obtained
7.52 $\pm$ 0.1 kpc for $M_{\rm K} = -1.59 \pm 0.03$, which translates
to 7.33 $\pm$ 0.14 kpc with the new DM. 

This can be compared to a distance of 7.94 $\pm$ 0.37 (random) $\pm$ 0.26 (systematic) kpc 
from Groenewegen et al. (2007) based on $K$-band monitoring observations of Population-{\sc ii} 
cepheids and RR Lyrae stars (distance tied to an assumed LMC distance modulus (DM) of 18.50) and
the direct determination of 7.73 $\pm$ 0.32 kpc based on the orbit of
the star called S2 around the central black hole when post-Newtonian
physics is taken into account (Eisenhauer et al. 2005, Zucker et
al. 2006), although neglect of the space motion of Sgr A$^*$ may
result in systematic errors of about 0.1-0.45 kpc (Nikiforov 2008).
All these independent observation are consistent and seem to indicate a
DM to the GC of slightly less than 8 kpc.

\acknowledgements{  
This research has made use of the SIMBAD database, operated at CDS,  Strasbourg, France. 
}

{}

\end{document}